\documentclass[nofootinbib,onecolumn,preprintnumbers,amsmath,amssymb,prd]{revtex4}

\begin{document}
\title{Generalized Lovelock gravity}
\preprint{ IPM/P-2009/047}

\author{Qasem \surname{Exirifard}}
\affiliation{School of Physics, Institute for Studies in Theoretical Physics and Mathematics (IPM), P.O.Box 19395-5531, Tehran, Iran}

\email{exir@theory.ipm.ac.ir}

\begin{abstract}
In the Riemann geometry, the metric's equation of motion for an arbitrary Lagrangian  is succinctly  expressed in term of the first variation of the action with respect to the Riemann tensor if the Riemann tensor were independent of the metric. Let this variation be called  the E-tensor. 

Noting that the E-tensor and equations of the motion for a general Lovelock gravity have the same differential degree, we define generalized Lovelock gravity  as polynomial scalar densities constructed out from the Riemann tensor and its arbitrary covariant derivatives such that they lead to the same differential degree for  the E-tensor and the metric's equation of motion.

We consider Lagrangian densities which are functional of the metric and the first covariant derivative of the Riemann tensor: ${\cal L}(\nabla_{a} R_{ijkl},g_{ij})$. We then present the first non-trivial examples of the generalized Lovelock gravity terms.
\end{abstract}
\maketitle

Quantization of gravity requires the Einstein-Hilbert action to be corrected. But what are the form of the corrections?   In string theory, corrections in the form of the covariant derivative of Riemann tensor exist in the sub-sub-leading levels \cite{Jack,srting1,srting2}. In the Horava-Lifshitz proposal  for quantum gravity in $3+1$ dimensions \cite{Horava}, the quantum corrections are functional of the covariant derivative of the Riemann tensor. So quantum corrections in $d=4$ most probably will depend also on the covariant derivatives of the Riemann tensor.\footnote{In the  phenomenological geometric approach, the dark energy and dark matter problems are evaluated as needs to modify the dynamics of the space-time in very low space-time curvature. The simplest geometric models to resolve these  \cite{Capozziello:2006uv,Capozziello:2006ph, Boehmer:2007kx} rely in the f(R) gravity \cite{DeFelice:2010aj}. However these models do not reproduce the Tully-Fisher relation \cite{Tully} and they  introduce a parameter that depends on the (baryonic) mass of the galaxy in order to account for its observed anomalous rotational velocity curve. Inclusion of the covariant derivatives of the Riemann tensor into the action  resolves these problems \cite{Exirifard:2008dy}. So studying actions that depends on the covariant derivatives of the Riemann tensor is admirable even within the phenomenological geometric approach to dark paradigms. } This dependency,  however,  increases the number of possibilities for the Lagrangian density: the number of algebraically independent polynomial scalar densities  constructed out from the Riemann tensor and  its covariant derivatives are larger than the number of those constructed out from  the Riemann tensor alone.  Thus it seems  plausible to provide a  criterion in order to systematically and self-consistently selects  a subset of the Lagrangians densities from all possibilities, a subset that also includes the covariant derivatives of the Riemann tensor. In this note, we aim to provide such a criterion.

A natural rephrasing of this aim  is  what the generalizations of the Einstein-Hilbert equation are. David Lovelock long time ago  posed this question. He demanded the generalized Einstein tensor to be: 
\begin{enumerate}
 \item  divergence free. 
 \item symmetric.
 \item concomitant of the first and second derivative of the metric. 
\end{enumerate}
In a set of publications \cite{Lovelock,Lovelock2,Lovelock-Actions}, he then  proved that the only Lagrangians meeting his demands are 
\begin{eqnarray}\label{Lovelock-action}
S_{\text{gravity}} &=&  \int d^D{x}\,\sqrt{-\det g}\,{\cal L}[g_{ij}, R_{ijkl}]\,,\\
{\cal L}[g_{ij}, R_{ijkl}] &=& \sum a_n {\cal L}_n[g_{ij}, R_{ijkl}]\,,\\
{\cal L}_n[g_{ij}, R_{ijkl}] & =& \theta^{i_1\cdots i_{2n}j_1 \cdots j_{2n}} \prod_{p=1}^{n} R_{i_{2p-1}i_{2p}j_{2p-1}j_{2p}}\,,
\end{eqnarray}
where
\begin{eqnarray}\label{delta}
\theta^{ \mu_1\cdots \mu_{2n} \nu_1 \cdots \nu_{2n}}(g) &=&\det
\left|
\begin{array}{lll}
g^{\mu_1 \nu_1} & \cdots & g^{\mu_{2n} \nu_1}\\
\vdots&&\vdots\\
g^{\mu_1 \nu_{2n}}&\cdots& g^{\mu_{2n} \nu_{2n}}
\end{array}\right|\,,
\end{eqnarray}
where $[\frac{D}{2}]$ represents the integer part of $\frac{D}{2}$, and $a_n$'s are some constant values. Neither of the above Lagrangians includes the covariant derivative of the Riemann tensor. So we do need to rephrase and modify the Lovelock criteria such that further possibilities are allowed. 

Having noted that ``a characteristic feature of Lovelock terms is that their first non vanishing term in the expansion of the metric around flat space-time is a total derivative, S. Cnockaert and M. Henneaux have investigated generalized Lovelock terms defined as polynomial scalar densities in the Riemann curvature tensor and its covariant derivatives (of arbitrarily high but finite order) such that their first non vanishing term in the expansion of the metric around flat space is a total derivative '' \cite{Cnockaert:2005jw}. They however have reported that their generalized Lovelock terms contains only the usual ones.

Naresh Dadhich  has provided a new independent identification of the Lovelock gravity from the Bianchi derivative of a curvature polynomial \cite{Dadhich:2008df}. One can investigate if this classification leads to a non-trivial generalization of Lovelock gravity. Ref. \cite{Exirifard:2007da,Borunda:2008kf} investigate and report the consistency\footnote{This is a strong form of the consistency, in the sense that all the solutions of the metric formulation are also the solutions of the first order (Palatini) formulation. Ref.\cite{Exirifard:2007da} shows these two formulations are equivalent  in  asymptotically `flat' space-time geometries.} of the Palatini (first order)\cite{Palatini-Form,Einstein,Error} formulation and the metric (second) formulation of the Lovelock terms. Ref.\cite{Exirifard:2007da}, having evaluated the consistency of the first and second order formulations as a criterion to restrict the form of the Lagrangian promises to apply this criterion on other Lagrangians. This promise is  yet to be  fulfilled.  

In this note we would like to look at a  characteristic of the Lovelock gravity that so far has  been overlooked in all attempts to generalize Lovelock gravity terms. To illustrate this characteristic let us look at the Gauss-Bonnet term which is $n=2$ term in \eqref{Lovelock-action} 
\begin{equation}\label{GBterm}
L_{GB} \,=\, R^{ijkl} R_{ijkl} - 4 R_{ij} R^{ij} + R^2 \,,
\end{equation}
The equation of motion for each of $R^{ijkl} R_{ijkl}$, $R_{ij} R^{ij}$ and $R^2$ is a fourth order differential equation. But the Guass-Bonnet term is the combination of these  terms that brings down the degree of differential equation of motion by two. Let us examine if the degree of the equation of motion can be brought down by two for a general action. In so doing we need the equation of motion for a general action.

A general action for the metric of a $D$ dimensional space-time can be presented as follows
\begin{eqnarray}
S_{\text{gravity}} &=&  \int d^D{x}\,\sqrt{-\det g}\,{\boldsymbol L}[g_{ij}, R_{ijkl},\nabla_i]\,,\\
{\boldsymbol L}[g_{ij}, R_{ijkl},\nabla_i] &=& {\cal L}[g_{ij}, R_{ijkl}, \nabla_a R^{ijkl}, \nabla_{(a_1}\nabla_{a_2)} R^{ijkl},\cdots,\nabla_{(a_1}\cdots\nabla_{a_n)}R^{ijkl} ]\,,
\end{eqnarray}
where $n$ is a natural number. The first variation of the action with respect to the metric then gives the equation of motion for the metric
\begin{eqnarray}
 -T^{ij} &=&\frac{\partial \cal L}{\partial g_{ij}} + \frac{1}{2} {\cal L}\, g^{ij} + E^{i}_{~\alpha \beta \gamma} R^{j \alpha \beta \gamma} + 2 \nabla_\alpha\nabla_\beta E^{i \alpha \beta j}\,,\label{EqMetric}\\
E^{ijkl} &=& \frac{\partial \cal L}{\partial R_{ijkl}} - \nabla_a \frac{\partial \cal L}{\partial \nabla_a R_{ijkl}} +\cdots + (-1)^n \nabla_{(a_1}\cdots \nabla_{a_n)} \frac{\partial \cal L}{\partial  \nabla_{(a_1}\cdots\nabla_{a_n)} R_{ijkl} }\,,\label{ETensor}
\end{eqnarray}
where $T^{ij}$ is the energy momentum tensor of matter fields minimally coupled to gravity, and partial derivatives of $L$ are taken assuming that $g_{ij}, R_{ijkl}, \cdots,\nabla_{(a_1}\cdots\nabla_{a_n)} R_{ijkl} $ are independent variables, and partial derivative coefficients are uniquely fixed to have precisely the same tensor symmetries as the varied quantities \cite{Iyer:1994ys}. The E-tensor\footnote{ Ref. \cite{Padmanabhan:2009jb} uses $P$ to refer to this tensor.  We use the notation of \cite{Iyer:1994ys} and call it the E-tensor.} \eqref{ETensor} would be the first variation of the action with respect to the Riemann tensor if the Riemann tensor were independent of the metric.

In the equation of motion \eqref{EqMetric}, $\nabla_\alpha\nabla_\beta E^{i \alpha \beta j}$  generally leads to  the appearance of two extra derivatives of the metric which are absent in the first three terms of the r.h.s of \eqref{EqMetric}.  In other words  the differential degree of the equation of motion is generally two degrees higher than that of the E-tensor.  Requiring the same differential degree for the E-tensor  and the equations of motion, thus, is a criterion to single out a specific set of Lagrangian from a larger given set.  We refer to this criterion as the E-criterion.\footnote{In order to have the Einstein-Hilbert action included, we should have rephrased the E-criterion as terms  for which the E-tensor's  degree is not larger than the degree of the equations of motion . We are, however, interested in the higher derivative corrections/modifications. So requiring the same degree for the equations of motion and the E-tensor suffices for our purpose.} The chosen/restricted subset has the privilege of leading to a differential equation of a lower degree than that of a general Lagrangian. Lovelock gravity terms are examples of these terms. So we name terms that satisfy the E-criterion as the generalized  Lovelock gravity terms. In the next section we consider Lagrangians which are functional of only the first covariant derivative of the Riemann tensor. We obtain a sufficient condition for the E-criterion. We find a family of solutions for this condition.  We then discuss on the uniqueness of the these terms. At the end, we will provide the summary and outlooks.

\section{Functional of the first covariant derivative of the Riemann tensor: ${\cal L}[g_{ij}, \nabla_a R_{ijkl}]$}
A Lagrangian in the form of ${\cal L}[g_{ij}, \nabla_a R_{ijkl}]$, polynomial in terms of $\nabla_a R_{ijkl}$ is a summation of the following terms
\begin{equation}\label{Lag}
{\cal L}_n \,=\, {\cal C}^{a_1 \cdots a_n i_1 \cdots i_{2n} j_1\cdots j_{2n}}~ \nabla_{a_1} R_{i_1i_2j_1j_2}\cdots \nabla_{a_n} R_{i_{2n-1}i_{2n}j_{2n-1} j_{2n}} 
\end{equation}
where ${\cal C}^{a_1 \cdots a_n i_1 \cdots i_{2n} j_1\cdots j_{2n}}$ is a functional of the metric components, and $n$ is a natural  number. Since  ${\cal C}^{a_1 \cdots a_n i_1 \cdots i_{2n} j_1\cdots j_{2n}}$ is a functional of only the metric's components it holds
\begin{equation}
 \nabla_p {\cal C}^{a_1 \cdots a_n i_1 \cdots i_{2n} j_1\cdots j_{2n}} \,=\,0\,.
\end{equation}
and  it  carries an even number of indices. Therefore $n$ is a natural even number.

We note that only the part of  ${\cal C}^{a_1 \cdots a_n i_1 \cdots i_{2n} j_1\cdots j_{2n}}$ which is symmetric under the exchange of $(a_m,i_{2m-1}, i_{2m},j_{2m-1},j_{2m})$ with $(a_p, i_{2p-1}, i_{2p},j_{2p-1}, j_{2p})$ for all $m$ and $p$ contributes to the Lagrangian density. So we choose ${\cal C}^{a_1 \cdots a_n i_1 \cdots i_{2n} j_1\cdots j_{2n}}$ such that it satisfies
\begin{eqnarray}\label{Sym-Acction}
&\,&{\cal C}^{a_1\cdots a_p\cdots a_m \cdots a_n i_1 \cdots i_{2p-1}i_{2p}\cdots i_{2m-1}i_{2m}\cdots i_{2n}j_1\cdots j_{2p-1} j_{2p}\cdots j_{2m-1}j_{2m}\cdots j_{2n}}\nonumber\\
&~~&= {\cal C}^{a_1\cdots a_m\cdots a_p \cdots a_n i_1 \cdots i_{2m-1}i_{2m}\cdots i_{2p-1}i_{2p}\cdots i_{2n}j_1\cdots j_{2m-1} j_{2m}\cdots j_{2p-1}j_{2p}\cdots j_{2n}}\,.
\end{eqnarray}
Let it be recalled that all the components of the Riemann tensor and its covariant derivatives are not algebraically independent. The Riemann tensor constructed out from the Levi Cevita connection satisfies:
\begin{eqnarray}
R_{i_1 i_2 j_1 j_2} + R_{i_2 i_1 j_1 j_2} &=&0\,,\label{R1}\\
R_{i_1 i_2 j_1 j_2} + R_{i_1 i_2 j_2 j_1} &=&0\,,\label{R2}\\
R_{i_1 i_2 j_1 j_2} -R_{ j_1 j_2 i_1 i_2} &=&0\,,\label{R3}\\
\nabla_{a_1} R_{i_1 i_2 j_1 j_2} + \nabla_{j_1} R_{i_1 i_2 j_2 a_1} + \nabla_{j_2} R_{i_1 i_2 a_1 j_1} &=& 0 \label{R4}
\end{eqnarray}
Because the Lagrangian density is the multiplication of the Riemann tensor and the ${\cal C}$-tensor, it  can be deduced from \eqref{R1}, \eqref{R2}, and \eqref{R3} that only the part of the ${\cal C}$-tensor that owns the following properties contributes to the Lagrangian:
\begin{subequations}\label{sys-C}
\begin{equation}
{\cal C}^{a_1\cdots a_n i_2 i_1  \cdots i_{2n}j_1\cdots j_{2n}} \,=\, -{\cal C}^{a_1\cdots a_n i_1 i_2  \cdots i_{2n}j_1\cdots j_{2n}}\,,
\end{equation}
\begin{equation}
{\cal C}^{a_1\cdots a_n i_1 \cdots i_{2n}j_2 j_1\cdots j_{2n}} \,=\, -{\cal C}^{a_1\cdots a_n i_1 \cdots i_{2n}j_1 j_2\cdots j_{2n}}\,,
\end{equation}
\begin{equation}
{\cal C}^{a_1\cdots a_n j_1 j_2 i_3\cdots i_{2n}i_1 i_2 j_3\cdots j_{2n}} \,=\, {\cal C}^{a_1\cdots a_n i_1  \cdots i_{2n}j_1\cdots j_{2n}}\,.
\end{equation}
\end{subequations}
The above relations  are nothing more than saying that when  an scalar is constructed out from the direct multiplication of a symmetric tensor and  a general tensor, then only the symmetric part of the general tensor contributes to the scalar.    We choose ${\cal C}$ such that it explicitly holds \eqref{sys-C}.

Now let it be noticed  that \eqref{R4} implies 
\begin{equation}
{\cal C}^{a_1 \cdots a_n i_1 \cdots i_{2n} j_1 \cdots j_{2n}}(\nabla_{a_1} R_{i_1 i_2 j_1 j_2} + \nabla_{j_1} R_{i_1 i_2 j_2 a_1} + \nabla_{j_2} R_{i_1 i_2 a_1 j_1}) = 0\,, 
\end{equation}
rewriting which yields
\begin{equation}\label{DR1=0}
({\cal C}^{a_1 a_2\cdots a_n i_1 i_2 i_3 \cdots i_{2n}j_1\cdots j_{2n}} + {\cal C}^{i_1 a_2\cdots a_n i_2 a_1 i_3 \cdots i_{2n}j_1\cdots j_{2n}}+{\cal C}^{i_2 a_2\cdots a_n a_1 i_1 i_3 \cdots i_{2n}j_1\cdots j_{2n}}\,) \nabla_{a_1} R_{i_1 i_2 j_1 j_2} =0\,,
\end{equation}
which  implies that  only the part of the ${\cal C}$-tensor that holds
\begin{equation}\label{DR=0}
{\cal C}^{a_1 a_2\cdots a_n i_1 i_2 i_3 \cdots i_{2n}j_1\cdots j_{2n}} + {\cal C}^{i_1 a_2\cdots a_n i_2 a_1 i_3 \cdots i_{2n}j_1\cdots j_{2n}}+{\cal C}^{i_2 a_2\cdots a_n a_1 i_1 i_3 \cdots i_{2n}j_1\cdots j_{2n}}\, =0\,,
\end{equation}
 contributes to \eqref{Lag}. In order to further clarify this statement, let the $\cal C$-tensor be re-expressed as follows
 \begin{eqnarray}\label{C-reex}
{\cal C}^{a_1 a_2\cdots a_n i_1 i_2 i_3 \cdots i_{2n}j_1\cdots j_{2n}} &=& {\hat C}^{a_1 a_2\cdots a_n i_1 i_2 i_3 \cdots i_{2n}j_1\cdots j_{2n}}+ {\cal A}^{a_1 a_2\cdots a_n i_1 i_2 i_3 \cdots i_{2n}j_1\cdots j_{2n}}  \,,\\
3{\cal A}^{a_1 a_2\cdots a_n i_1 i_2 i_3 \cdots i_{2n}j_1\cdots j_{2n}} &=& {\cal C}^{a_1 a_2\cdots a_n i_1 i_2 i_3 \cdots i_{2n}j_1\cdots j_{2n}} + {\cal C}^{i_1 a_2\cdots a_n i_2 a_1 i_3 \cdots i_{2n}j_1\cdots j_{2n}}+{\cal C}^{i_2 a_2\cdots a_n a_1 i_1 i_3 \cdots i_{2n}j_1\cdots j_{2n}}\,. 
 \end{eqnarray}
 Then when we insert \eqref{C-reex} into the Lagrangian density, due to \eqref{DR1=0}, we see that only the ${\hat C}^{a_1 a_2\cdots a_n i_1 i_2 i_3 \cdots i_{2n}j_1\cdots j_{2n}}$ - the part of the $\cal C$-tensor that holds \eqref{DR=0} - contributes to the Lagrangian density.  So we  choose the ${\cal C}$-tensor such that it meets  \eqref{DR=0}. Note that neither this choice nor  \eqref{Sym-Acction},  nor \eqref{sys-C} affects the generality of the considered Lagrangians.

Eq. \eqref{sys-C}  and \eqref{DR=0} indicates that the ${\cal C}$-tensor carries all the symmetries of the covariant derivative of the Riemann tensor. So the partial derivative of the ${\cal L}_n$ with respect to $\nabla_{a_1} R_{ijkl}$ can be simply written by
\begin{equation}
\frac{\partial {\cal L}_n}{\partial \nabla_a R_{i_1 i_2 j_1 j_2}} \,=\, n {\cal C}^{a_1 a_2\cdots a_n i_1 i_2 i_3 \cdots i_{2n}j_1\cdots j_{2n}} \prod_{p=2}^{n}\nabla_{a_p} R_{i_{2p-1}i_{2p}j_{2p-1}j_{2p}}\,.
\end{equation}
If we had not chosen the ${\cal C}$-tensor to satisfy \eqref{Sym-Acction}, \eqref{sys-C}  and \eqref{DR=0}, then $\frac{\partial {\cal L}_n}{\partial \nabla_a R_{i_1 i_2 j_1 j_2}}$ could not been written in the above compact form. The E-tensor can be written in a compact form too:
\begin{equation}\label{E-com}
E_{{\cal L}_n}^{i_1 i_2 j_1 j_2}\,=\,-\nabla_{a_1}\frac{\partial {\cal L}_n}{\partial \nabla_a R_{i_1 i_2 j_1 j_2}} \,=\, - n(n-1) {\cal C}^{a_1 a_2\cdots a_n i_1 i_2 i_3 \cdots i_{2n}j_1\cdots j_{2n}} \nabla_{a_1}\nabla_{a_2} R_{i_3i_4 j_3 j_4}\prod_{p=3}^{n}\nabla_{a_p} R_{i_{2p-1}i_{2p}j_{2p-1}j_{2p}}\,.
\end{equation}
Note that the E-tensor is a functional of the Riemann tensor and its first two covariant derivatives. The equation of motion for ${\cal L}_n$ also can be simplified to 
\begin{equation}\label{EqLn}
\left((\frac{\partial}{\partial g_{ab}} + \frac{1}{2} g^{ab}) {\cal C}^{a_1\cdots a_n i_1 \cdots i_{2n} j_1 \cdots j_{2n}}\right)\prod_{p=1}^{n}\nabla_{a_p} R_{i_{2p-1}i_{2p}j_{2p-1}j_{2p}}  + E^{a \alpha \beta \gamma}_{{\cal L}_n} R^{b}_{~\alpha\beta\gamma} + 2 \nabla_{\alpha}\nabla_{\beta} E^{a\alpha\beta b}_{{\cal L}_n} \,=\,0\,,
\end{equation}
for which the E-criterion requires $\nabla_{\alpha}\nabla_{\beta} E^{a\alpha\beta b}_{{\cal L}_n}$ to be a functional of the Riemann tensor and its first two derivatives. From \eqref{E-com},   $\nabla_{i_2}\nabla_{j_1} E^{i_1\alpha\beta j_2}_{{\cal L}_n}$ follows
\begin{equation}
\nabla_{i_2}\nabla_{j_1} E^{i_1i_2 j_1 j_2}_{{\cal L}_n}\,=\,-\nabla_{i_2}\nabla_{j_1}\nabla_{a_1}\frac{\partial {\cal L}_n}{\partial \nabla_a R_{i_1 i_2 j_1 j_2}}\,
\end{equation}
which can be re-expressed by
\begin{equation}\label{EDLDR}
\nabla_{i_2}\nabla_{j_1} E^{i_1i_2 j_1 j_2}_{{\cal L}_n}\,=\,-(\nabla_{i_2} [\nabla_{j_1},\nabla_{a_1}]+ [\nabla_{i_2}, \nabla_{a_1}] \nabla_{j_1}+ \nabla_{a_1} \nabla_{i_2} \nabla_{j_1})\frac{\partial {\cal L}_n}{\partial \nabla_a R_{i_1 i_2 j_1 j_2}}\,,
\end{equation}
Noting that commutators can be expressed in terms of the Riemann tensor, we conclude  $\nabla_{i_2}\nabla_{j_1} E^{i_1\alpha\beta j_2}_{{\cal L}_n}$ and $\nabla_{a_1} \nabla_{j_1} \nabla_{i_2}\frac{\partial {\cal L}_n}{\partial \nabla_a R_{i_1 i_2 j_1 j_2}}$ have the same differential degree. So applying the E-criterion on ${\cal L}_n$ is the same as requiring $\nabla_{a_1} \nabla_{j_1} \nabla_{i_2}\frac{\partial {\cal L}_n}{\partial \nabla_a R_{i_1 i_2 j_1 j_2}}$ to be a functional of the Riemann tensor and its first two covariant derivatives.

Now let us look at  $\nabla_{i_2} \nabla_{j_1}\frac{\partial {\cal L}_n}{\partial \nabla_{a_1} R_{i_1 i_2 j_1 j_2}}$:
\begin{equation}
\nabla_{i_2} \nabla_{j_1}\frac{\partial {\cal L}_n}{\partial \nabla_a R_{i_1 i_2 j_1 j_2}} \,=\, n {\cal C}^{a_1 a_2\cdots a_n i_1 i_2 i_3 \cdots i_{2n}j_1\cdots j_{2n}}  \nabla_{i_2}\nabla_{j_1} \left( \prod_{p=2}^{n}\nabla_{a_p} R_{i_{2p-1}i_{2p}j_{2p-1}j_{2p}}\right)\,,
\end{equation}
expanding which yields
\begin{eqnarray}\label{i2j1DLDR}
\nabla_{i_2} \nabla_{j_1}\frac{\partial {\cal L}_n}{\partial \nabla_a R_{i_1 i_2 j_1 j_2}} &=& n (n-1) {\cal C}^{a_1 a_2\cdots a_n i_1 i_2 i_3 \cdots i_{2n}j_1\cdots j_{2n}}  \nabla_{i_2} \nabla_{j_1}\nabla_{a_2}   R_{i_3 i_4 j_3 j_4} \prod_{p=3}^{n}\nabla_{a_p} R_{i_{2p-1}i_{2p}j_{2p-1}j_{2p}}\,+\nonumber\\
&&+ n(n-1)(n-2) 
{\cal C}^{a_1 a_2\cdots a_n i_1 i_2 i_3 \cdots i_{2n}j_1\cdots j_{2n}} 
\nabla_{j_1}\nabla_{a_2} R_{i_3 i_4 j_3 j_4}
\nabla_{i_2}\nabla_{a_3} R_{i_5 i_6 j_5 j_6}\times\nonumber\\
&& \qquad \qquad \qquad
\prod_{p=4}^{n}\nabla_{a_p} R_{i_{2p-1}i_{2p}j_{2p-1}j_{2p}}\,.
\end{eqnarray}
Now suppose that it holds 
\begin{equation}\label{Chold}
{\cal C}^{a_1 a_2\cdots a_n i_1 i_2 i_3 \cdots i_{2n}j_1\cdots j_{2n}} \nabla_{i_p} R_{i_{2p-1} i_{2p} j_{2p-1} j_{2p}} \,=\,0\,,
\end{equation}
then\footnote{Commutators are used to have $\nabla_{i_2}$ and  $\nabla_{j_1}$ act on the Riemann tensor before $\nabla_{a_i}$.}
\begin{subequations}\label{whatlabeltochooseforlthis}
\begin{eqnarray}
{\cal C}^{a_1 a_2\cdots a_n i_1 i_2 i_3 \cdots i_{2n}j_1\cdots j_{2n}} \nabla_{j_1} \nabla_{a_2} R_{i_3 i_4 j_3 j_4}&=& {\cal C}^{a_1 a_2\cdots a_n i_1 i_2 i_3 \cdots i_{2n}j_1\cdots j_{2n}} [\nabla_{j_1}, \nabla_{a_2}] R_{i_3 i_4 j_3 j_4} \,,\\
{\cal C}^{a_1 a_2\cdots a_n i_1 i_2 i_3 \cdots i_{2n}j_1\cdots j_{2n}}  \nabla_{j_2} \nabla_{a_3} R_{i_5 i_6 j_5 j_6}&=& {\cal C}^{a_1 a_2\cdots a_n i_1 i_2 i_3 \cdots i_{2n}j_1\cdots j_{2n}} [\nabla_{i_2}, \nabla_{a_2}] R_{i_5 i_6 j_5 j_6}\,, \\
{\cal C}^{a_1 a_2\cdots a_n i_1 i_2 i_3 \cdots i_{2n}j_1\cdots j_{2n}}\nabla_{i_2} \nabla_{j_1} \nabla_{a_2} R_{i_3 i_4 j_3 j_4}&=& {\cal C}^{a_1 a_2\cdots a_n i_1 i_2 i_3 \cdots i_{2n}j_1\cdots j_{2n}} [\nabla_{i_2},[\nabla_{j_1}, \nabla_{a_2}]] R_{i_3 i_4 j_3 j_4}\,.
\end{eqnarray}
\end{subequations}
Inserting \eqref{whatlabeltochooseforlthis}  into \eqref{i2j1DLDR} yields
\begin{eqnarray}\label{i2j1DLDR-1}
\nabla_{i_2} \nabla_{j_1}\frac{\partial {\cal L}_n}{\partial \nabla_a R_{i_1 i_2 j_1 j_2}} &=& n (n-1) {\cal C}^{a_1 a_2\cdots a_n i_1 i_2 i_3 \cdots i_{2n}j_1\cdots j_{2n}}  
\left([\nabla_{i_2},[\nabla_{j_1}, \nabla_{a_2}]] R_{i_3 i_4 j_3 j_4}\right)
 \prod_{p=3}^{n}\nabla_{a_p} R_{i_{2p-1}i_{2p}j_{2p-1}j_{2p}}\,+\nonumber\\
&&+ n(n-1)(n-2) 
{\cal C}^{a_1 a_2\cdots a_n i_1 i_2 i_3 \cdots i_{2n}j_1\cdots j_{2n}} 
\left([\nabla_{j_1}, \nabla_{a_2}] R_{i_3 i_4 j_3 j_4})\right)
\left( [\nabla_{i_2}, \nabla_{a_2}] R_{i_5 i_6 j_5 j_6}\right)
\times\nonumber\\
&& \qquad \qquad \qquad
\prod_{p=4}^{n}\nabla_{a_p} R_{i_{2p-1}i_{2p}j_{2p-1}j_{2p}}\,.
\end{eqnarray}
\eqref{i2j1DLDR-1} proves that when \eqref{Chold} holds then $\nabla_{i_2} \nabla_{j_1}\frac{\partial {\cal L}_n}{\partial \nabla_a R_{i_1 i_2 j_1 j_2}}$ is a functional of the Riemann tensor and its first covariant derivative. So  when \eqref{Chold} holds  $\nabla_{a_1}\nabla_{i_2} \nabla_{j_1}\frac{\partial {\cal L}_n}{\partial \nabla_a R_{i_1 i_2 j_1 j_2}}$ is a functional of the Riemann tensor and its first \textit{two} covariant derivatives: the E-criterion for ${\cal L}_n$ is satisfied.  Recalling $\nabla_{[i_1} R_{i_2 i_3] j_2 j_3}=0 $,  \eqref{Chold} is guaranteed to be satisfied if 
\begin{equation}\label{TheCondition2e}
 {\cal C}^{a_1\cdots a_n {\bold i_1 \bold i_2  \bold i_3} \cdots i_{2n} j_1  \cdots j_{2n}}
 \,=\,{\cal C}^{a_1\cdots a_n {\bold i_2 \bold i_3  \bold i_1} \cdots i_{2n} j_1  \cdots j_{2n}}
 \,=\,{\cal C}^{a_1\cdots a_n {\bold i_3 \bold i_1  \bold i_2} \cdots i_{2n} j_1  \cdots j_{2n}}\,,
 \end{equation}
Let it be highlighted  that \eqref{TheCondition2e} is a sufficient condition for the $\cal C$-tensor to meet the E-criterion.

\subsection{Finding the first generalized Lovelock gravity terms}
The $\theta$ tensor \eqref{delta} yields
\begin{eqnarray}
 {\theta}^{ i_1 i_2 i_3 \cdots i_{2n} j_1\cdots j_{2n}}\nabla_{i_p} R_{i_{2p-1} i_{2p} j_{2p-1} j_{2p}} \,=\,0\,,
 \\
 {\theta}^{{\bold i_1 \bold i_2  \bold i_3} \cdots i_{2n} j_1  \cdots j_{2n}}
 \,=\,{\theta }^{{\bold i_2 \bold i_3  \bold i_1} \cdots i_{2n} j_1  \cdots j_{2n}}
 \,=\,{\theta}^{ {\bold i_3 \bold i_1  \bold i_2} \cdots i_{2n} j_1  \cdots j_{2n}}\,,
\end{eqnarray}
which forgetting the $a_1\cdots a_n$ are similar to \eqref{Chold} and \eqref{TheCondition2e}. This similarity motivates us to consider the following Lagrangian as an ansatz that meets the E-criterion: 
\begin{equation}\label{ansatz}
{\cal L}_n \,=\, \tilde{{\cal C}}^{a_1 \cdots a_n} \theta^{i_1 \cdots i_{2n} j_1\cdots j_{2n}}~ \nabla_{a_1} R_{i_1i_2j_1j_2}\cdots \nabla_{a_n} R_{i_{2n-1}i_{2n}j_{2n-1} j_{2n}} \,,
\end{equation}
where $\theta$ is given in \eqref{delta} and $\tilde{{\cal C}}^{a_1 \cdots a_n}$ can be chosen to be  symmetric under exchange of each of its two indices. 

We note that  though $\tilde{{\cal C}}^{a_1 \cdots a_n} \theta^{i_1 \cdots i_{2n} j_1\cdots j_{2n}}$ yields \eqref{Chold} and \eqref{TheCondition2e}:
\begin{equation}\label{TDR=0}
 {\tilde C}^{a_1\cdots a_n} \theta^{{\bold i_1 \bold i_2  \bold i_3} \cdots i_{2n} j_1  \cdots j_{2n}}
 \,=\,{\tilde C}^{a_1\cdots a_n}\theta^{ {\bold i_2 \bold i_3  \bold i_1} \cdots i_{2n} j_1  \cdots j_{2n}}
 \,=\,{\tilde C}^{a_1\cdots a_n} \theta^{{\bold i_3 \bold i_1  \bold i_2} \cdots i_{2n} j_1  \cdots j_{2n}}\,,
\end{equation}
it does not have all the symmetries of the $\cal C$-tensor: it does not hold \eqref{DR=0}. We should pay attention that \eqref{TheCondition2e} is a sufficient condition for the E-criterion provided that the coefficients of the covariant derivative of the Riemann tensor hold all the required symmetries of the C-tensor. 

Let us decompose  $\tilde{{\cal C}}^{a_1 \cdots a_n} \theta^{i_1 \cdots i_{2n} j_1\cdots j_{2n}}$ to the part that meets  \eqref{Sym-Acction}, \eqref{sys-C}  and \eqref{DR=0}, and the rest:
\begin{equation}\label{Atensor}
\tilde{{\cal C}}^{a_1 \cdots a_n} \theta^{i_1 \cdots i_{2n} j_1\cdots j_{2n}} = \hat{C}^{a_1 \cdots a_n i_1 \cdots i_{2n} j_1\cdots j_{2n}} + A^{a_1 \cdots a_n i_1 \cdots i_{2n} j_1\cdots j_{2n}} \,,
\end{equation}
where   $\hat{C}^{a_1 \cdots a_n i_1 \cdots i_{2n} j_1\cdots j_{2n}}$ respects \eqref{Sym-Acction}, \eqref{sys-C}  and \eqref{DR=0}, while $A^{a_1 \cdots a_n i_1 \cdots i_{2n} j_1\cdots j_{2n}}$ does not respect these symmetries. It is worth noting that only the $\hat{C}^{a_1 \cdots a_n i_1 \cdots i_{2n} j_1\cdots j_{2n}}$ contributes to the Lagrangian density
\begin{eqnarray}
{\cal L}_n &=& \tilde{{\cal C}}^{a_1 \cdots a_n} \theta^{i_1 \cdots i_{2n} j_1\cdots j_{2n}}~ \nabla_{a_1} R_{i_1i_2j_1j_2}\cdots \nabla_{a_n} R_{i_{2n-1}i_{2n}j_{2n-1} j_{2n}}\\
&=& \hat{C}^{a_1 \cdots a_n i_1 \cdots i_{2n} j_1\cdots j_{2n}}~ \nabla_{a_1} R_{i_1i_2j_1j_2}\cdots \nabla_{a_n} R_{i_{2n-1}i_{2n}j_{2n-1} j_{2n}}\,,\nonumber
\end{eqnarray}
Lagrangian density is independent of $A^{a_1 \cdots a_n i_1 \cdots i_{2n} j_1\cdots j_{2n}}$.
It then follows from \eqref{TDR=0} and \eqref{Atensor} that the $\hat{C}^{a_1 \cdots a_n i_1 \cdots i_{2n} j_1\cdots j_{2n}}$ meets \eqref{TheCondition2e}.\footnote{Note that also $A^{a_1 \cdots a_n i_1 \cdots i_{2n} j_1\cdots j_{2n}}$ has the symmetries of \eqref{TheCondition2e}.} So \eqref{ansatz} meets the E-criterion and indeed is a generalized Lovelock Gravity.
 
\section{on the uniqueness of the found generalized lovelock gravity terms}

In this section we wish to test/prove  uniqueness of some special cases of \eqref{ansatz}. Let us consider the simplest choice of ${\cal L}_n$, the first example of \eqref{ansatz}  for $n=2$:
\begin{equation}\label{GGB}
{\cal L}_{2} \,=\, \nabla_a R_{ijkl} \nabla^a R^{ijkl} - 4 \nabla_a R_{ij} \nabla^a R^{ij} + \nabla_a R \nabla^a R\,. 
\end{equation}
which should be called the generalization of the Gauss-Bonnet Lagrangian density.\footnote{ In contrary to the Gauss-Bonnet Lagrangian \eqref{GGB} is not a topological term in $D=4$. This  can easily be verified by evaluating the action of ${\cal L}_{2}$ for :
\begin{equation*}\label{sphericalmetric}
ds^2 \,=\, -A(r) dt^2 + \frac{ dr^2}{A(r)} + r^2 (d\theta^2 + \sin^2 \theta d\phi^2)\,.
\end{equation*}
The ${\cal L}_{2}$ action evaluated for  \eqref{sphericalmetric} reads: 
\begin{equation*}
S_{{\cal L}_2} = 4\pi \int dt \int dr \frac{4 A}{r^3}((-4A' + 4A'' r-2A''' r^2) (A-1) +r^3 A''^2 + 3 A'^2  + r^3 A''' A' - 4 r^2 A'' A'  )\,,
\end{equation*}
which is not a total derivative or a topological term:  $\frac{\delta S_{{\cal L}_2}}{\delta A} \neq 0$. 
}$^{,}$\footnote{The explicit form for $n=4$ can be written using the compact form of the forth order Lovelock gravity that is presented in \cite{fourth}.} 
Perhaps \eqref{GGB} is not the most general Lagrangian quadratic in term of the first covariant derivative of the Riemann tensor. But is it the only  one satisfying the E-criterion?

 In the following, we  shall prove that \eqref{GGB} is the only Lagrangian quadratic in term of the first covariant derivative of the Riemann tensor meeting the E-criterion. In so doing we notice that  the algebraically independent scalars which are quadratic in the Riemann tensor and quadratic in the  covariant derivative read \cite{Fulling}
\begin{eqnarray}
 R \Box R, ~ \nabla^p \nabla^q R R_{pq},\nabla_s \nabla_r R_{pq} R^{prqs},~R^{pq}\Box R_{pq}\,\\ \nonumber
 \nabla^p R \nabla_p R,~\nabla^r R^{pq} \nabla_r R_{pq},\nabla_rR^{pq} \nabla_q R_{pr},~\nabla^t R^{pqrs} \nabla_t R_{pqrs}\,.
\end{eqnarray}
The most general Lagrangian density which is quadratic both in the Riemann tensor and the covariant derivative is a linear combination of all the above possibilities
\begin{eqnarray}\label{LagrangianDR20}
L &=& c_1 R \Box R + c_2 \nabla^p \nabla^q R R_{pq} + c_3 \nabla_r \nabla_s R_{pq} R^{prqs} + c_4 R^{pq} \Box R_{pq} + \\
  && + c_5 \nabla^p R  \nabla_p R + c_6 \nabla^r R^{pq} \nabla_r R_{pq} + c_7 \nabla^r R^{pq} \nabla_q R_{pr} + c_8 \nabla^t R^{pqrs} \nabla_t R_{pqrs}\,, \nonumber
\end{eqnarray} 
where $c_1\cdots c_8$ are some constants real values. Performing integration by parts and using  
\begin{eqnarray}
\nabla^p R_{pq}&=&\frac{1}{2} \nabla_q R\,,\label{Rij=R}\\
\nabla_s R^{prqs}&= &\nabla^p R^{qr}-\nabla^r R^{qp}\,,
\end{eqnarray}
the Lagrangian density \eqref{LagrangianDR20} can be rewritten to
\begin{equation}\label{LagrangianDR21}
L \,=\, (c_5 -\frac{c_2}{2}-c_1) \nabla^p R \nabla_p R + (c_6-c_4-c_3) \nabla_r R_{pq} \nabla^{r} R^{pq} - (c_7-c_3) R_{pq} \nabla_r \nabla^p R^{qr}+ c_8 \nabla_t R_{pqrs}\nabla^t R^{pqrs}\,.
\end{equation} 
Noticing the algebraic identity of 
\begin{equation}
 \nabla_r \nabla^p R^{qr} \,=\, [\nabla_r,\nabla^p] R^{qr} + \nabla^p \nabla_r R^{qr} \,=\, [\nabla_r,\nabla^p] R^{qr}+ \frac{1}{2} \nabla^p \nabla^q R\,, 
\end{equation} 
and performing an integration by part, and using \eqref{Rij=R},   \eqref{LagrangianDR21} can be re-expressed by 
\begin{equation}\label{LagrangianDR21-2}
L \,=\, (c_5 -\frac{c_2}{2}-c_1+\frac{c_7-c_3}{4}) \nabla^p R \nabla_p R + (c_6-c_4-c_3) \nabla_r R_{pq} \nabla^{r} R^{pq} + c_8 \nabla_t R_{pqrs}\nabla^t R^{pqrs}- \,(c_7-c_3) R_{pq}[\nabla_r,\nabla^p] R^{qr}\,.
\end{equation} 
The commutators of the covariant derivatives can be expressed in term of the Riemann tensor. So $R_{pq}[\nabla_r,\nabla^p] R^{qr}$ is   cubic in term of the Riemann tensor.\footnote{This term relies within the family of the Lagrangians which are functional of the Riemann tensor not its covariant derivative. In this family of the Lagrangian only the ordinary Lovelock gravity meets the E-criterion.} We are interested in  actions which are quadratic in term of the covariant derivative, we thus set $c_7=c_3$. We also redefine the constants values
\begin{eqnarray}
a_1 & = & c_5 -\frac{c_2}{2} -c_1\,,\\
a_2 & = & c_6 - c_4 - c_3\,,\\
a_3 & = & c_8\,.
\end{eqnarray}
The most general Lagrangian density which is quadratic both in the Riemann tensor and the covariant derivatives, therefore, reads
\begin{equation}\label{LagrangianDR22}
L \,=\, a_1 \nabla^p R \nabla_p R + a_2 \nabla_r R_{pq} \nabla^{r} R^{pq} + a_3 \nabla_t R_{pqrs}\nabla^t R^{pqrs}\,.
\end{equation} 
For  general values of $a_1, a_2$ and $a_3$, \eqref{LagrangianDR22} leads to six order differential equations. We would like to find all values of $a_1$, $a_2$ and $a_3$ for which \eqref{LagrangianDR22} leads to  fourth order equations (imposing the E-criterion on \eqref{LagrangianDR22} is the same as requiring it to lead to fourth order equations). Since ${\cal L}_2$ \eqref{GGB} leads to  fourth order equations, and equations of motion are linear in term of the Lagrangian density, subtracting a multiplication of ${\cal L}_2$ from \eqref{LagrangianDR22} does not change the differential degree of the equations of motion derived from  \eqref{LagrangianDR22}. We do the following subtraction:
\begin{equation}\label{LagrangianDR23}
\tilde{L} \,=\, L - a_3 {\cal L}_2 \,= \, (a_1-a_3) \nabla^p R \nabla_p R + (a_2 +4 a_3) \nabla_r R_{pq} \nabla^{r} R^{pq}\,. 
\end{equation} 
Any values of $a_1, a_2$ and $a_3$ that leads to  fourth order equations  derived from \eqref{LagrangianDR22}, leads to  fourth  order equations derived from \eqref{LagrangianDR23} and vice versa. We define the following constant values in order to write \eqref{LagrangianDR23} in a more compact form:
\begin{eqnarray}\label{ba-realation}
b_1 & = & a_1 - a_3 \,,\\
b_2 & = & a_2 + 4 a_3 \,.\nonumber 
\end{eqnarray}
Using the above constant values, $\tilde{L}$ reads
\begin{equation}\label{LagrangianDR24}
\tilde{L} \,=\, b_1 \nabla^p R \nabla_p R + b_2 \nabla_r R_{pq} \nabla^{r} R^{pq}\,. 
\end{equation} 
We will prove that only for $b_1=b_2=0$, $\tilde{L}$ leads to  fourth order equations.  Then $b_1=b_2=0$  implies that ${\cal L}_2$ (up to an overall factor) is the most general Lagrangian density, quadratic in term of the Riemann tensor and quadratic in the covariant derivative leading to   fourth order equations for the metric's components.   

Instead of considering a general metric and calculating the functional variation of  \eqref{LagrangianDR23}, let us consider a general time-independent spherical metric in $d=4$: 
\begin{eqnarray}\label{metric-test}
ds^2 \,=\, -A(r) dt^2 + \frac{dr^2}{A(r) B(r)} + r^2 (d\theta^2+ \sin^2 \theta d\phi^2)\,,
\end{eqnarray}
and calculate the functional variation of \eqref{LagrangianDR24} with respect to $A(r)$ and $B(r)$. It is known that the functional variation does not generally commute with imposing symmetries on the solution; here we have imposed spherical symmetry and time translation. We notice, however, our imposed symmetries are isometries of the Riemann manifold -supposedly a smooth manifold- so the principle of symmetric criticality is met  \cite{Palais}. In other words the functional  variation of the  action corresponding to \eqref{LagrangianDR24} computed for \eqref{metric-test},  indeed gives the equations of motion of $A(r)$ and $B(r)$.  We will show that only for $b_1=b_2=0$, $\tilde{L}$ leads to  fourth order equations for $A(r)$ and $B(r)$. So $b_1=b_2=0$ must hold in order to have fourth order equation for a general metric. 

In the following we are going to compute the equations of motion for $A(r)$ and $B(r)$ in \eqref{metric-test}. The  action corresponding to \eqref{LagrangianDR24} computed for  \eqref{metric-test} reads 
\begin{equation}\label{SLfinale}
S_{\tilde L}\,=\, 4 \,\pi \int dt~\int dr\,  r^2\,\sqrt{\frac{1}{B}}(b1 \left.\nabla_p R \nabla^p R\right|_{\text{Eq}. \eqref{metric-test} }+ b_2 \left.\nabla_r R_{pq} \nabla^r R^{pq}\right|_{\text{Eq}. \eqref{metric-test}}  )\,,
\end{equation}
where 
\begin{eqnarray}
\frac{4\, r^6\,\left. \nabla_a R \nabla^a R\right|_{\text{Eq}. \eqref{metric-test}} }{A\, B}&=& \left( 
-  B'' A'r^3
- 3 B' A'' r^3
 - 12 B' A' r^2
- 2 B A'''r^3\right.\\
&&\left.
~
- 8B A''r^2
+ 4Br A'
- 4A B''r^2 
- 8 + 
8\,A\,B\right)^2\,, \nonumber\\
\frac{8\,r^6 \left.\nabla_r R_{pq} \nabla^r R^{pq}\right|_{\text{Eq}. \eqref{metric-test}} }{A\, B}&=&96
+ 80\,r^4\,B'\, A''\,A'\,B
+ 12\,r^6\, B'\,A''\,B\, A'''
- 32\,r^3\,B^2\, A''\,A'
+ 8\,r^5\, B''\,A'\,B\, A'' 
\\ \nonumber &&
+ 4\,r^6\,B''\, A'\,B\,A'''
+ 64\,A\,B'\,r
+ 96\,B\,A''\,r^2
+ 24\,r^5\,B'\,A'\,B\,A'''
- 48\,r^3\, B'\,A'^2\,B 
\\ \nonumber&&
+ 12\,r^5\,B''\, A'^2\,B'
- 8\,r^4\, B'\,A\,B\,A'''
+ 52\,r^4\,B'\,A\,B''\,A'
- 16\,r^3\,A'\,B\,A\,B'' 
\\\nonumber&&
- 8\,r^4\,B''\, A'^2\,B
- 192\,A\,B
+ 12\,r^5\,B'\,A''\,A\, B''
+ 8\,r^5\,B\, A'''\,A\,B'' 
\\ \nonumber&&
+ 32\,r^4\,B\,A''\,A\,B''
- 96\,r^2\,B'\,A\,A'\,B
+ 16\,r^3\,B\, A''\,B'\,A
+ 36\,r^5\, B'^2\,A''\, A' 
\\ \nonumber&&
+ 6\,r^6\,B''\, A'\,B'\,A''
+ 16\,r^5\,B^2\,A'''\,A''
+ 24\,r^5\,B'\,A''^2\,B
+ 32\,A\, B''\,r^2 
\\ \nonumber&&
+ 112\,B'\,A'\,r^2
+ 96\,B^2\,A^2
- 32\,A^2\,B''\,r^2\,B
- 64\,A^2\,B'\,r\,B
- 96\,B^2\,A''\,r^2\,A 
\\ \nonumber&&
+ 4\,r^5\,B''^2\, A'\,A
- 8\,r^3\,B'\,A^2\, B''
- 12\,r^4\,B'^2\,A\,A''
- 16\,r^4\,B^2\,A'''\,A' 
\\ \nonumber&&
 + 12\,r^4\,A^2\,B''^2
+ 20\,r^2\,B'^2\,A^2
+ 40\,r^4\,B^2\,A''^2
+ 4\,r^6\,B^2\,A'''^2
+ 9\,r^6\,B'^2\,A''^2 
\\ \nonumber&&
+ r^6\,B''^2\,A'^2
+ 16\,r^2\,A'^2\,B^2
+ 82\,B'^2\,A'^2\,r^4\,.
\end{eqnarray}
The functional variations of \eqref{SLfinale} with respect to $A(r)$ and $B(r)$ at most are functional of the first six derivatives of $A(r)$ and $B(r)$, and $r$. It is straightforward to calculate them. It is then easy to show that  
\begin{eqnarray}\label{DLDA1}
\label{DLDA4}
\frac{\partial}{\partial B^{(5)}(r)}\left(\frac{\delta S_{\tilde L}}{\delta A}\right)  &=& \frac{r B^{\frac{3}{2}}}{2}(2 (4 b_1 + b_2) A + (2 b_1 + b_2) r A')\,,
\end{eqnarray} 
	If the equations of motion of $A$ and $B$ are fourth order, then \eqref{DLDA4} must algebraically vanish. It algebraically vanishes only for 
\begin{equation}\label{b1b2=0}
b_1 \,=\, b_2 \,=\, 0 \,.
\end{equation}
Since there exists no non-vanishing value for $b_1$ and $b_2$  leading to fourth order equations for spherical time-independent metric, then there exists no non-vanishing value for  $b_1$ and $b_2$ that leads to fourth order equations for components of a general metric. To put it in other words, \eqref{b1b2=0} besides \eqref{ba-realation} proves that ${\cal L}_2$ \eqref{GGB} is the only Lagrangian in the form of \eqref{LagrangianDR22} that satisfies the E-criterion.   We expect that such a uniqueness property could be generalized to also include other ${\cal L}_n$: ${\cal L}_n$ is the only action constructed from the multiplication of $n$-times of the first covariant derivative of the Riemann tensor which satisfies the E-criterion. In other words, we expect it to be proved that \eqref{TheCondition2e} is also the necessary condition for the E-criterion, and \eqref{ansatz} is its only solution.

\section{Summary and outlooks}
The differential degree for the equations of motion of metric  generally is two degrees higher than that of the first variation of the Lagrangian with respect to the Riemann tensor if the Riemann tensor were independent of metric (the E-tensor). So requiring the same differential degree for the  metric's equation and the E-tensor serves as a criterion to single out a subset of Lagrangians from a given larger family of Lagrangian. This criterion (the E-criterion) can be utilized when one wishes to fix the field  redefinition ambiguities \cite{Tseytlin} in quantum loop corrections. 

We have noticed that Lovelock gravity is the result of applying our criterion to Lagrangians which are functional of the metric and the Riemann tensor. We have considered Lagrangians in the form of ${\cal L}(g_{ij},\nabla_a R_{ijkl})$ polynomial in term of the first covariant derivative of the metric. We have found that \eqref{ansatz} meets the criterion. In particular in $d=4$, \eqref{GGB} lead to fourth order differential equation for the metric. So it should be considered when one addresses a general fourth order gravity \cite{Schmidt:2006jt}.

Perhaps we can apply the E-criterion on other families of Lagrangians: families  that includes arbitrary higher derivatives of the Riemann tensor. The E-criterion then chooses a subset of these Lagrangians.

Notice that our criterion is a weaker condition than that of requiring (a strong form of) consistency between Palatini and metric formulations \cite{Exirifard:2007da,Borunda:2008kf}. It, however, sounds interesting to check if  adding terms cubic in  the Riemann tensor to \eqref{GGB}, would lead to a Lagrangian whose  Palatini and metric formulations are (strongly) consistent. 

\providecommand{\href}[2]{#2}\begingroup\raggedright

\end{document}